\documentclass[letter]{spie}
\usepackage[]{graphicx,color,amsmath,amsfonts,amssymb}

\title{3D laser printing by ultra-short laser pulses for micro-optical  applications: towards telecom wavelengths}
\author{Meguya Ryu\supit{1}, Vygantas Mizeikis\supit{2}, Junko
Morikawa\supit{1}, Hernando Magallanes\supit{3},\\\protect Etienne
Brasselet\supit{3}, Simonas Varapnickas\supit{4}, Mangirdas
Malinauskas\supit{4}, Saulius Juodkazis\supit{5,6} \skiplinehalf
\supit{1}Tokyo Institute of Technology, Meguro-ku, Tokyo
152-8550, Japan\\
\supit{2}Research Institute of Electronics, Shizuoka University,
3-5-1 Johoku, Naka-ku, Hamamatsu 432-8561, Japan\\
\supit{3} Universit\'e Bordeaux, CNRS, LOMA, UMR5798,  351 Cours
de la Lib\'eration, 33405 Talence, France\\ \supit{4}Department of
Quantum Electronics, Physics Faculty, Vilnius University,
Saul\.{e}tekio Ave. 10, LT-10223, Vilnius,
Lithuania\\\supit{5}{Nanotechnology facility, Swinburne
Univerisity of Technology, John st., Hawthorn, 3122 Vic,
Australia} \\\supit{6}{Melbourne Centre for Nanofabrication, the
Victorian Node of the Australian National Fabrication Facility,
151 Wellington Rd., Clayton 3168 Vic, Australia} }
\authorinfo{Further author information: \\S.J.: E-mail:
sjuodkazis@swin.edu.au, Telephone: +61 3 9214 87178}
\begin{document}
\maketitle
\begin{abstract}
Three dimensional (3D) fast ($< 0.5$~hour) printing of
micro-optical elements down to sub-wavelength resolution over
$100~\mu$m footprint areas using femtosecond (fs-)laser oscillator
is presented. Using sub-1~nJ pulse energies, optical vortex
generators  made of polymerised grating segments with an
azimuthally changing orientation have been fabricated in SZ2080
resist; width of polymerised rods was $\sim 150$~nm and period
0.6-1~$\mu$m. Detailed phase retardance analysis was carried out
manually with Berek compensator (under a white light illumination)
and using an equivalent principle by an automated Abrio
implementation at 546~nm. Direct experimental measurements of
retardance was required since the period of the grating was
comparable (or larger) than the wavelength of visible light. By
gold sputtering, transmissive optical vortex generators were
turned into reflective ones with augmented retardance, $\Delta
n\times h$ defined by the form birefringence, $\Delta n$, and the
height $h = 2d$ where $d$ is the thickness of the polymerised
structure. Retardance reached $315$~nm as measured with Berek
compensator at visible wavelengths. Birefringent phase delays of
$\pi$ (or $\lambda/2$ in wavelength) required for high purity
vortex generators  can be made based on the proposed approach.
Optical vortex generators for telecom wavelengths with
sub-wavelength patterns of azimuthally oriented gratings are
amenable by direct laser polymerisation.
\end{abstract}


\section{INTRODUCTION}
\label{sec:intro}  

Micro-optical elements and photonic wire bonding in telecom
applications are becoming essential building blocks for imaging,
surveillance, telecommunications, security, optical fiber, and
sensor technologies~\cite{16le16133}. Simplification of processing
and fabrication steps is always recognisable in industrial
innovations. In 3D nano-/micro-printing during last decade, we
have seen emergence of new photo-materials tailored for laser
fabrication using ultra-short sub-ps laser pulses. Different
direct write, holographic exposure, nanoparticle-mediated modes of
material
modifications~\cite{06njp250,06apl221101,09jpc11720,09jpc1147,13oe6901},
formation of micro-channels in polymers and
glasses~\cite{04apa1549,03apa371,02ass705p}, using different beam
intensity profiles~\cite{06ol80,01jjap1197} have been tested. Many
remarkable results have been achieved with amplified femtosecond
(fs)-laser systems at lower laser repetition rates when thermal
effects, usually not desirable, can be avoided. However, even
simpler solutions are available using only fs-laser oscillators
with pulse energies of $\sim  1~$nJ at high repetition rate of
$\sim 80$~MHz and allow harnessing of photo-thermal effects which
delivers more control in 3D laser ablation~\cite{03apl2901} and
polymerisation. By fast beam scanning, it is possible to reach a
high sub-wavelength resolution and practical fabrication times of
micro-optical elements with sub-mm cross sections demonstrated in
this study and not possible to achieve with low repetition rate
fs-laser systems~\cite{02ass705p}.

Here, we show 3D fabrication of flat optical elements by using
only fs-laser oscillator over area of 0.1~mm in cross section.
High fidelity fabrication was achieved with uniform height of the
optical vortex generators chosen to belong to the family of the
so-called q-plates. The q-plates are birefringent patterns (here
form-birefringent) whose slow-optical axis has a local azimuthal
angular orientation $\theta = q\alpha$ defined by the azimuth
$\alpha$ with $q$ being
half-integer~\cite{Biener,Marrucci2006,Sergei}. They can be
realised by different fabrication and patterning
methods~\cite{Biener,Shimotsuma,Karimi,Guixin,Jin,Capasso2016,Kruk,Kamali}
and allow polarisation controlled management of the orbital
angular momentum. Birefringent phase retardance of $\pi$ has been
achieved (at visible wavelengths) for the reflection-type q-plates
using one layer fabrication. This fabrication is simpler as
compared with q-plates polymerised using direct writing with
amplified fs-laser system~\cite{17apl}. Optical characterisation
of form-birefringent flat optical elements was carried out to
inspect structural quality of patterns, which are sub-wavelength
for the telecommunication wavelengths where flat optical elements
for generation of optical vortex beams carrying orbital angular
momentum (OAM) are under active exploration due to possibility to
reach high data transfer densities.

\begin{figure}[tb!]
\centering
\includegraphics[width=\textwidth]{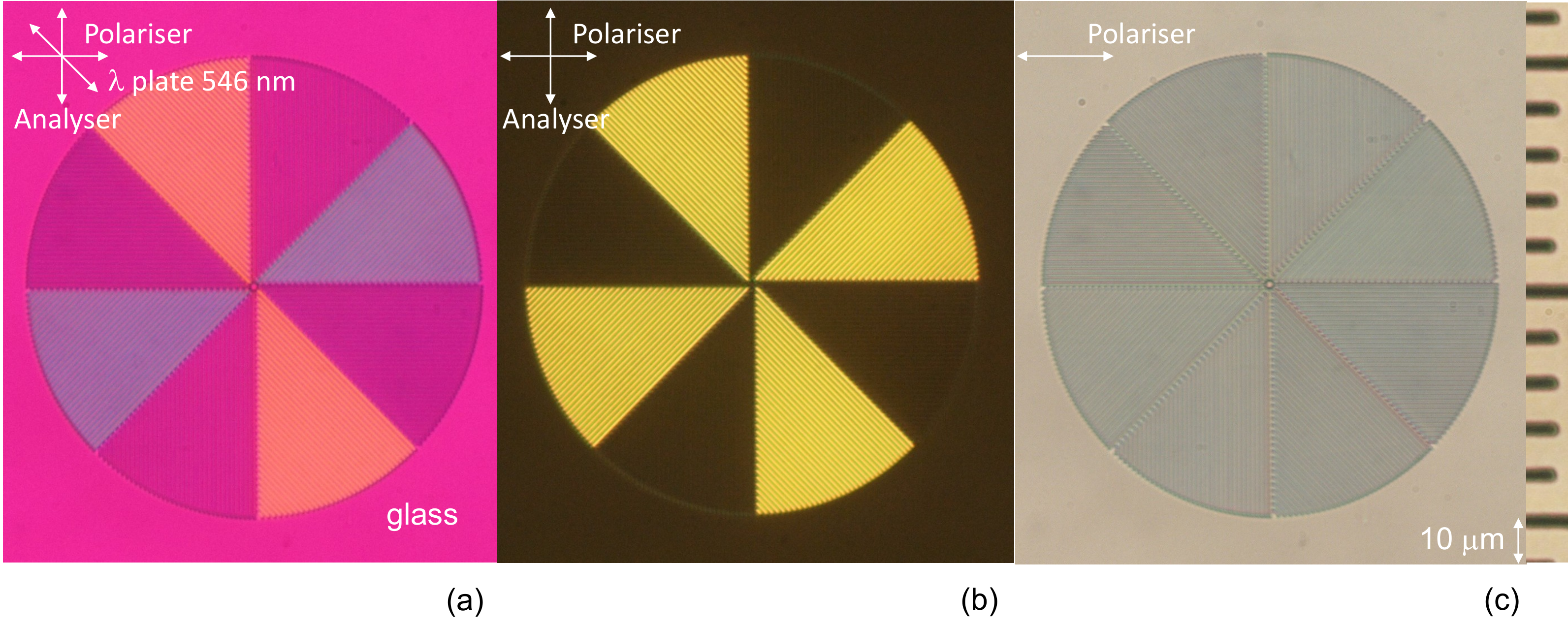}
\caption{(a) A cross-polarised wave-plate color-shifted image of a
3D laser polymerised q-plate ($q = 1$). (b) Cross-polarised image.
(c) Optical image with only polariser. Material: SZ2080 with 0.5\%
Bis photo-initiator, laser pulse energy $E_p = 0.15$~nJ (power
$P_{av} = 12$~mW at 80~MHz repetition rate), wavelength $\lambda =
800$~nm. } \label{f-opti}
\end{figure}

\section{Experimental}

\subsection{Fabrication of q-plates}

Mai Tai (Spectra Physics) fs-laser oscillator emitting 800~nm
wavelength, 120~fs duration pulses was used as a light source for
direct laser writing. The pulses were focused into the photoresist
through a microscope cover glass substrate using a microscope
objective lens (Olympus, UPlanSApo 60$^\times$/1.35 Oil) with
numerical aperture $NA = 1.35$. During the direct laser writing
the sample was scanned by a 3D piezo-stage (Physik Instrumente,
P-563.3CD) with xy-(in-plane) stroke of $300~\mu$m and z-axis
(axial) stroke of 250$\,\mu$m. Typical writing speed was
$20~\mu$m/s. High pulse repetition rate of 80~MHz ensured that
focal area of diameter $2w_0 =1.22\lambda/NA \simeq 0.72~\mu$m was
exposed to millions of laser pulses. Circularly polarised laser
beam was used in experiments in order to equalize lateral diameter
of polymerised voxels and obtain lines, whose width does not
depend on the orientation~\cite{16aom1209}. The initial focusing
plane for fabrication was chosen by adjusting the z-axis position
within $\leq 0.1~\mu$m accuracy window. The empirical procedure to
find glass substrate/photoresist interface  involved determination
of z-axis position at which brightness of two-photon excited
photoluminescence emitted by the photoinitiator reached half of
its maximal value.  The experimental system used for fabrication
ensured that z-axis position was maintained stable from within few
tens of minutes to hours, which provided sufficient time for
sample fabrication. However, random drifts of the z-axis position
by up to 1$\,\mu$m also occurred occasionally, which led to line
height and retardance variation across the area of q-plate, and
degraded optical quality of some samples. Precise origin of this
drift has not yet been determined, although it is likely related
to thermal deformation of metallic parts in the setup, and to
capillary drag of the moving sample by the oil-immersion lens.

\begin{figure}[tb!]
\centering
\includegraphics[width=\textwidth]{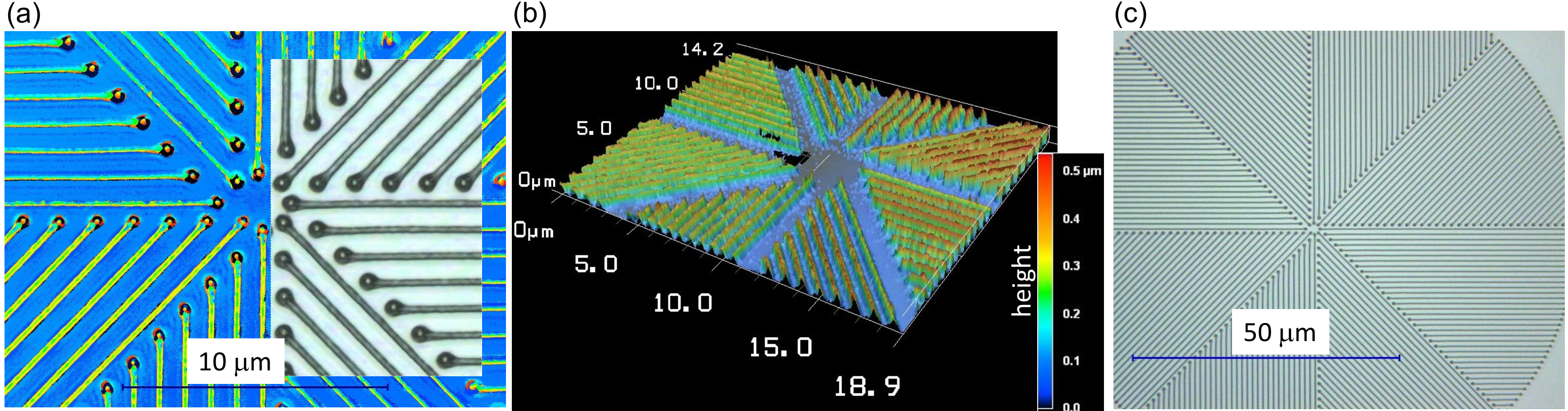}
\caption{(a) Optical profilometer image of a $q=1$ plate of
$\Lambda = 1~\mu$m period with an overlayed segment of an optical
image. (b) Height profile. (c) Optical image of larger area.
Material: SZ2080 with 0.5\% Irg. photo-initiator, laser pulse
energy $E_p = 0.31$~nJ (power $P_{av} = 25$~mW at 80~MHz
repetition rate), wavelength $\lambda = 800$~nm. } \label{f-high}
\end{figure}

The negative-tone Zr-containing hybrid organic-inorganic
photoresist SZ2080 with only 0.5\% of Irgacure 369
(2-Benzyl-2-dimethylamino-1-(4-morpholinophenyl)-butanone-1) and
(Irg) and Michler's ketone 4,4'-Bis (diethylamino) benzophenone
(Bis) was added as photo initiators. Low concentration of photo
initiator reduces an optical absorption at the visible spectral
range and, consequently, reduces dichroism of form-birefringent
q-plates which are inherently spectrally broad band optical
elements, though at the expense of wavelength-dependent
efficiency. The q-plates were prepared by drop-casting photoresist
on a microscope cover glass substrates (Matsunami) and
subsequently drying them on a hot plate using temperature ramp
(for 5~min) between 40, 60, and 80$^\circ$C for 20~min. After
laser exposure, the samples were developed in
1-propanol:isopropanol (50:50) solution for 5~min., rinsed in
ethanol, and dried in super-critical CO$_2$ using a critical-point
dryer (JCPD-5, JEOL), in order to eliminate destructive action of
capillary forces during conventional drying.

\subsection{Characterisation of q-plates}

Retardance of form-birefringent q-plates $\Delta n\times h$~[nm]
was measured using Berek compensator No. 10412 (Nichika, Co. Ltd.)
setup on a Nikon Optiphot-Pol microscope and by using manual
alignment; $\Delta n$ is the birefringence and $h$ is an optical
path length. Calcite compensator plate is inserted at 45$^\circ$
to the crossed analyser-polariser orientation along the slow axis
of the extraordinary refractive index $n_e$; calcite is the
negative uniaxial material $n_o > n_e$. Then, by tilting the
compensator plate clockwise and anti-clockwise, a selected and
aligned form-birefringent region of q-plate was made darkest (the
lowest transmission); the $n_e$ orientation along the polymerised
grating of the q-plate was aligned with $n_o$ orientation of the
Berek compensator. The average angle between two settings was used
to find the retardance using tabulated reference. Also, an Abrio
attachment to Nikon microscope was used to determine the
birefringence (retardance) and orientation of the optical
fast-axis at 546~nm wavelength. This measurement is carried out
automatically and was compared with manual measurements with Berek
compensator. These two different methods were applied in
collaborating labs in Bordeaux and Tokyo and calibrated using
commercial quarter-waveplates of the known $\pi/2$ phase
retardance.

\section{Results}

Figure~\ref{f-opti} shows different optical images of $q = 1$
plate which reveal a high quality and uniformity of the
polymerised structure. By inserting a 530~nm waveplate at
45$^\circ$ orientation into cross-polarised imaging setup, a color
shift helps to reveal even better settle changes in the phase
retardance. The close up view (Fig.~\ref{f-opti}(a)) shows the
darker regions where SZ2080 polymerised logs were fabricated and
color tint almost reaches that of the substrate's between the logs
(an air gap region). Hence, there is the same light phase at the
bottom between the logs and glass substrate surface. Such imaging
provides a detailed assessment of birefringent phase delay at
different locations of this form-birefringent structure and is
valuable due to limitations of the effective medium theory (EMT)
predictions which are only valid for patterns with period much
smaller than the wavelengths of the light $\Lambda\ll\lambda$. In
this particular case $\Lambda = 1~\mu$m while the $\lambda\simeq
530$~nm, obviously out of validity range of EMT.

\begin{figure}[tb!]
\centering
\includegraphics[width=0.55\textwidth]{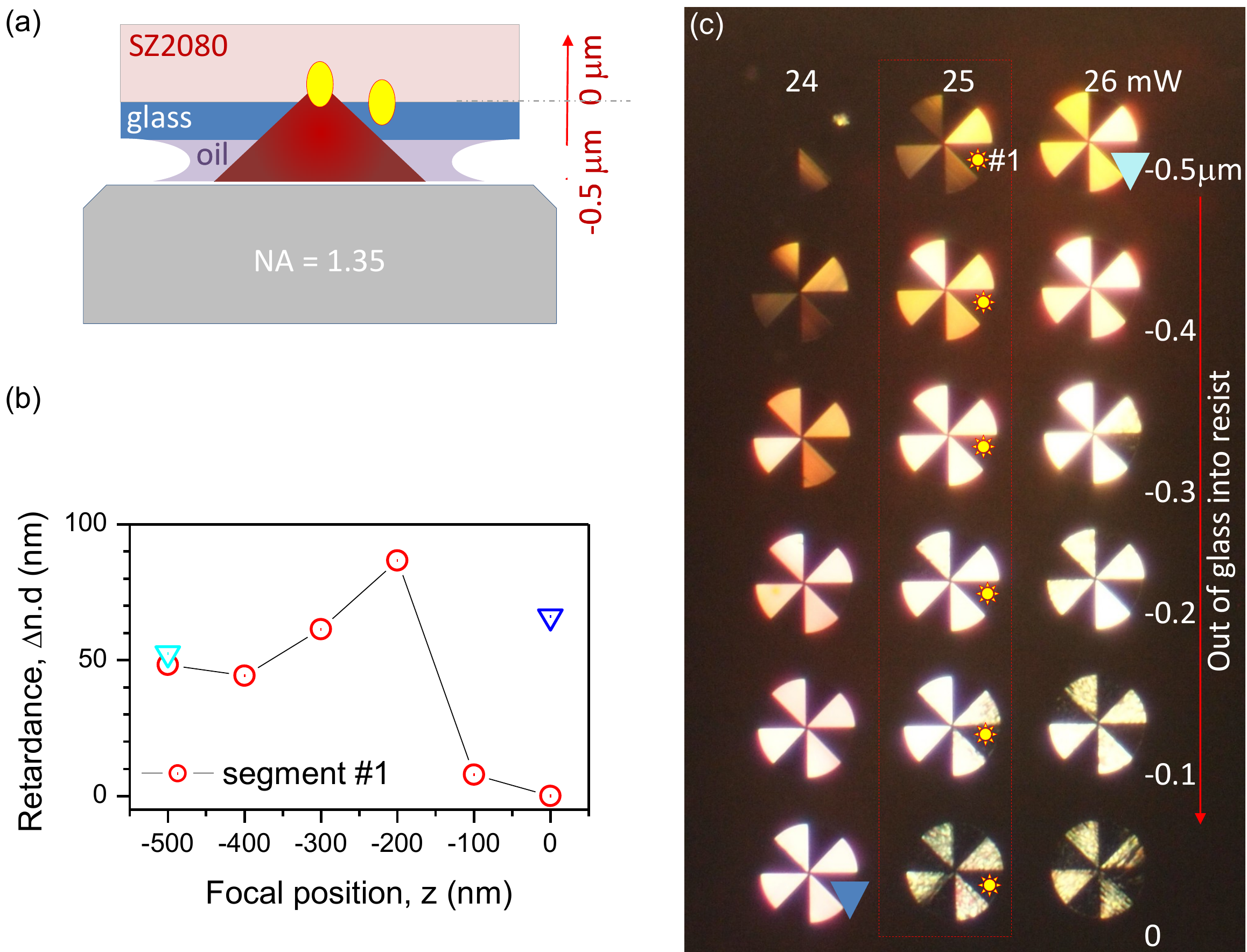}
\caption{(a) Schematic of focusing used to write $q=1$ plates. (b)
Dependence of the $\Lambda = 0.6~\mu$m period q-plates retardance
$\Delta n \times d$, where $d$ is height, $\Delta n$ is refractive
index change \emph{vs} focal position $z$; $z = 0$ was decided by
imaging of the interface between resist and glass. (c) Optical
cross-polarised image of fabricated structures. Retardance of
segment $\#1$ of the structures shown inside rectangular area was
measured with Berek compensator under white light condenser
illumination. Material: SZ2080 with 0.5\% Irg. photo-initiator,
laser pulse energy $E_p = 0.31$~nJ (power $P_{av} = 25$~mW at
80~MHz repetition rate), wavelength $\lambda = 800$~nm. }
\label{f-hight}
\end{figure}
\begin{figure}[tb!]
\centering
\includegraphics[width=0.65\textwidth]{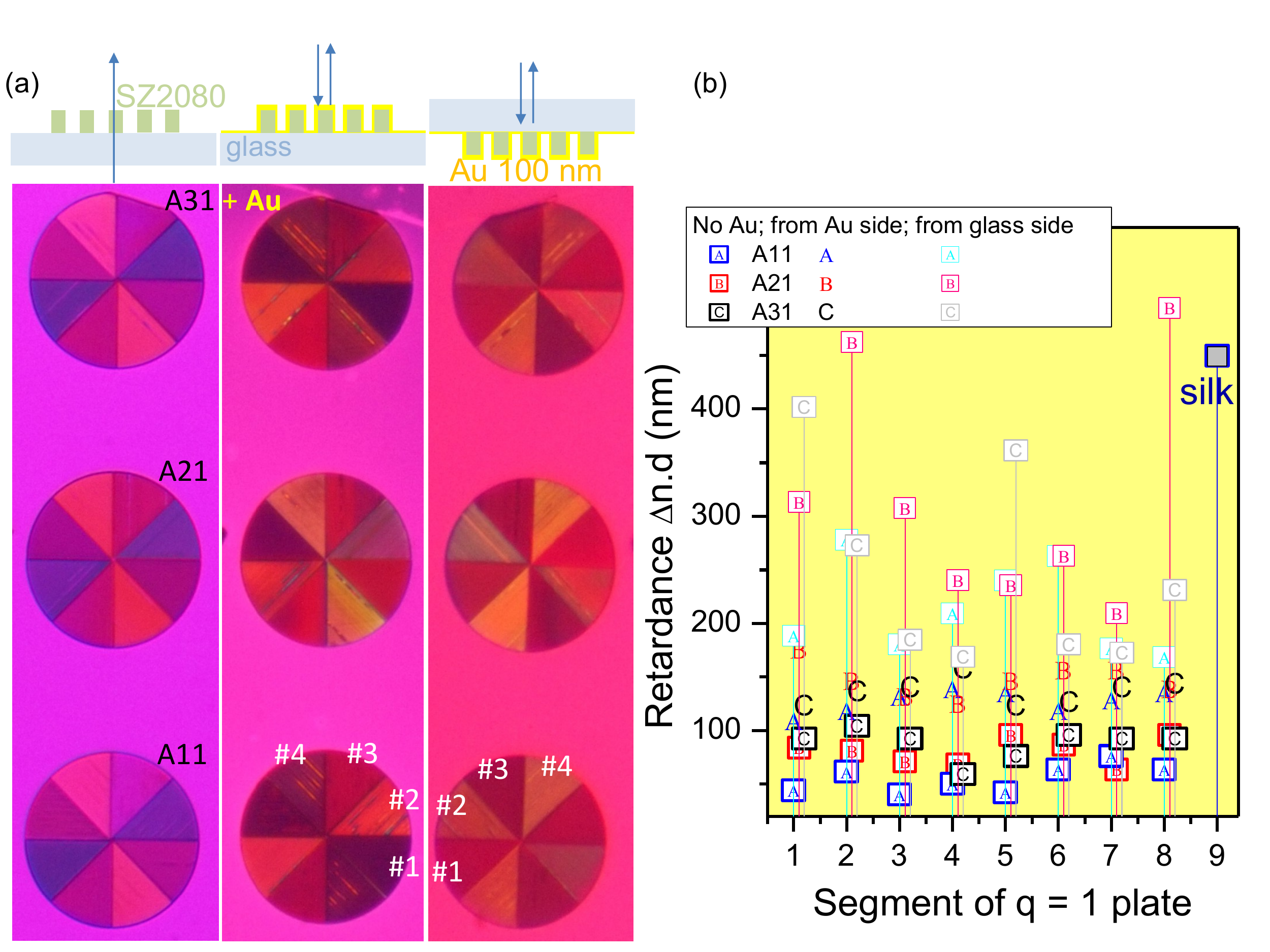}
\caption{(a) Color-shifted cross-polarised images of the same
$q=1$ plates as fabricated and then coated with 100~nm of gold and
imaged from gold and glass substrate sides, respectively (as shown
schematically on the top). (b) Retardance $\Delta n\times d$
measured manually on separate segments with Berek compensator.
Conditions of fabrication the same as for Fig.~\ref{f-opti}. }
\label{f-flip}
\end{figure}

Figure~\ref{f-high} provides optical profilometry of the q-plate
which has period $\Lambda = 1~\mu$m and height of $d \simeq
0.4~\mu$m for the width of polymerised logs $w \simeq 150$~nm.
Even high duty-cycle $w/\Lambda \simeq 0.5$ could be fabricated
using such conditions with air gap of 150~nm. The pattern over
0.1~mm-diameter was fabricated with high fidelity over a practical
time span of 30~min. The start and stop points were overexposed
what caused thickening of rods. This can be removed using a more
sophisticated shutter control which was not implemented in this
first fabrication.

Figure~\ref{f-hight} shows dependence of retardance as depth of
focusing was changed (along z-axis; see panel (a)). The structural
quality easily distinguishable from optical images correlated with
the retardance, which reached the highest values for the tallest
patterns (Fig.~\ref{f-hight}(b)). These q-plates
(Fig.~\ref{f-hight}(c)) have period of $\Lambda = 0.6~\mu$m and
duty cycle close to 0.5.

\begin{table}[t]
\caption{Average retardance of the same q-plates: as fabricated
and Au-coated (measured from two sides). Sample is shown in
Fig.~\ref{f-flip}. }
\begin{center}\label{table1}
 \begin{tabular}{||c c c c||}
 \hline
 Sample &                  Transmission           & Reflection from Au-side & Reflection from Au through substrate \\ [0.5ex]
 & $\Delta n\times h$(nm):                &  $\Delta n\times h$(nm):   & $\Delta n\times h$(nm): \\
 \hline\hline
 A11 & 55.3 & 127.8 & 213.0 \\
 \hline
 A21 & 80.9 & 148.0 & 315.7 \\
 \hline
 A31 & 88.0 & 135.9 & 246.7 \\
 \hline
\end{tabular}
\end{center}
\end{table}

\begin{figure}[b]
\centering
\includegraphics[width=0.5\textwidth]{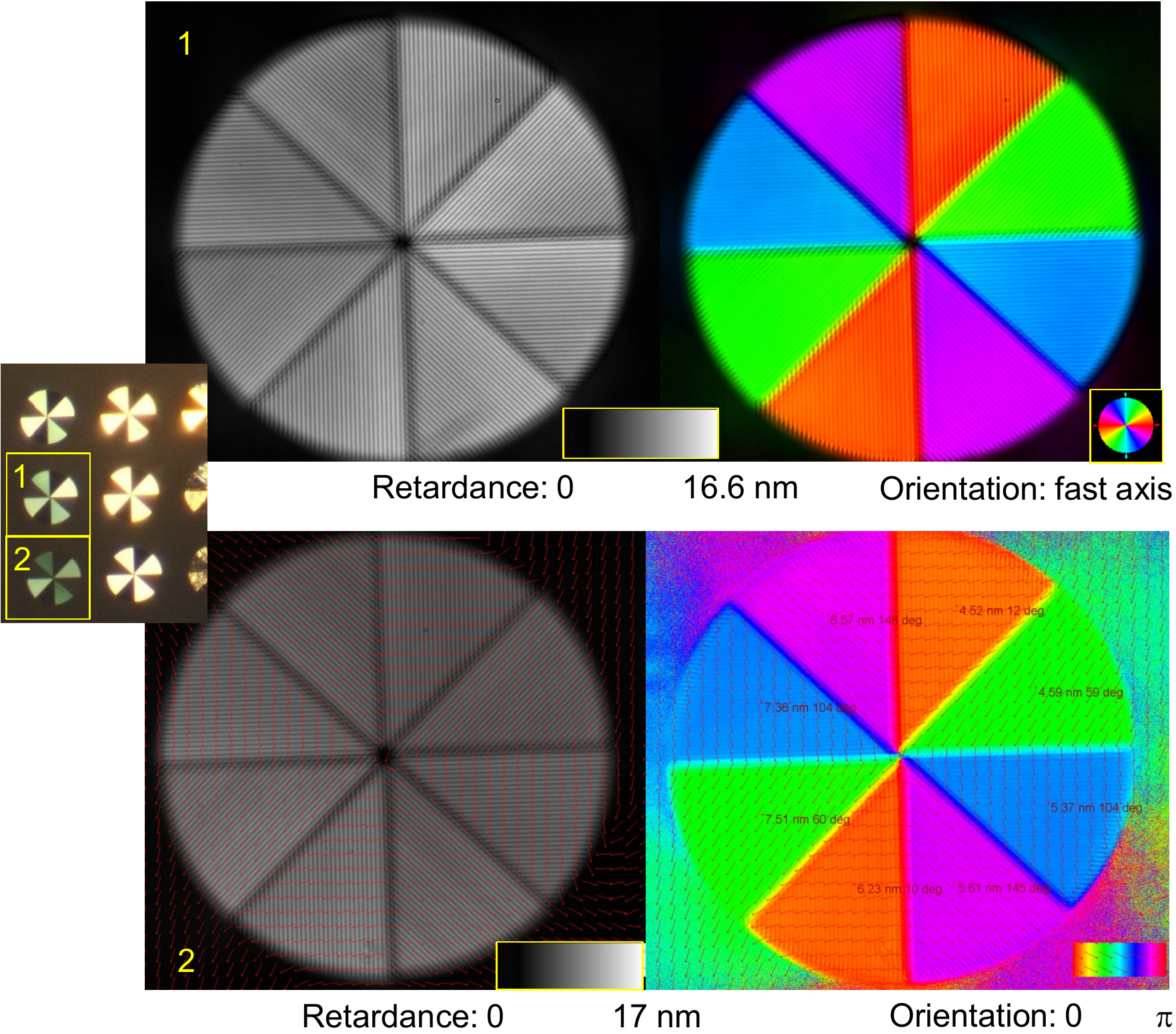}
\caption{Retardance and orientation of the optical fast; along
$n_e$ axis of $q =1$ plates measured in transmission by Abrio at
546~nm. Inset (left-middle) shows the q-plates polymerised at
slightly different pulse energies.  Two different presentations
data are chosen for the samples 1 and 2. Material: SZ2080 with
0.5\% Irg photo-initiator, laser pulse energy $E_p = 0.30$~nJ
(sample 1 at power $P_{av} = 24$~mW) and $E_p = 0.31$~nJ (sample 2
at $P_{av} = 25$~mW) for 80~MHz repetition rate, wavelength
$\lambda = 800$~nm.} \label{f-abrio}
\end{figure}

By evaporating 100~nm film of gold directly over q-plate, the
optical transmission is blocked. In reflection mode, there is a
benefit of doubling the optical path length $h = 2d$, which
increases retardance $\Delta n\times h$. Moreover, by measuring
retardance in reflection from the q-plate side this doubling take
place in air while by flipping the sample and measuring from the
glass side allows to double the optical path in SZ2080 portion of
the q-plate (see schematics on top of Fig.~\ref{f-flip}(a)). The
color-shifted cross-polarised images of the described above three
modes with different retardance are gathered in
Fig.~\ref{f-flip}(a).

Measurements of retardance with Berek compensator at each segment
(eight per q-plate) are summarised in Fig.~\ref{f-flip}(c) and the
average is presented in Table~\ref{table1}. Quite large data
scatter was observed. For comparison, retardance of a single silk
fiber was also measured (diameter $\sim 20~\mu$m) with
birefringence found $\Delta n \simeq 0.022\pm 0.002$, which is
typical to silk (Fig.~\ref{f-flip}(b)). Since the measurements
were carried out with Berek's compensator aligned along the silk
fiber (alignment along fast axis $n_o$ of the compensator) a
positive retardance is consistent with a larger refractive index
of silk for the E-field polarised along the fiber.

As fabricated q-plates had an average (over 8 segments) retardance
of 55.3 nm (A11), 80.9 nm (A21), and 88.0 nm (A31) summarised in
Table~\ref{table1}. When measured in reflection from the gold
deposited side, retardance increased by 2.31 for A11, by 1.83 for
A21, and by 1.54 for A31. The highest values were obtained for the
reflection measurements from the substrate side  with further
increase by 1.67 (A11), by 2.13 (A21), and by 1.82 (A31); the
increase is compared with the reflection case from the q-plate
side. These values are already larger than $\lambda/2$ or $\pi$ in
terms of the phase delay. These retardance increases are
approximately scaling as $n\times h$ ($n$ is the refractive
medium) and reach factor of $\sim 3$ for the reflection from the
substrate side as expected. Since $\Lambda \simeq \lambda$, this
experimental observation provides a direct measurement which
cannot be obtained by EMT modeling. Due to a manual nature of
measurements with Berek compensator, there is a considerable
scatter of results due to a visual judgement for the darkest
segments.

A simpler solution of reflective q-plate was realised by
Au-evaporation on the back-side of substrate (not on the q-plate
as discussed above). Doubling of retardance was achieved, however,
a strong hallo effects, i.e., a wide area of leaking illumination
through cross polariser-analyser setup was observed around the
q-plate area. This is caused by scattering and redirection of
light as it traverses twice through the entire thickness of cover
glass $\sim 200~\mu$m and q-plate.

\begin{figure}[tb!]
\centering
\includegraphics[width=0.85\textwidth]{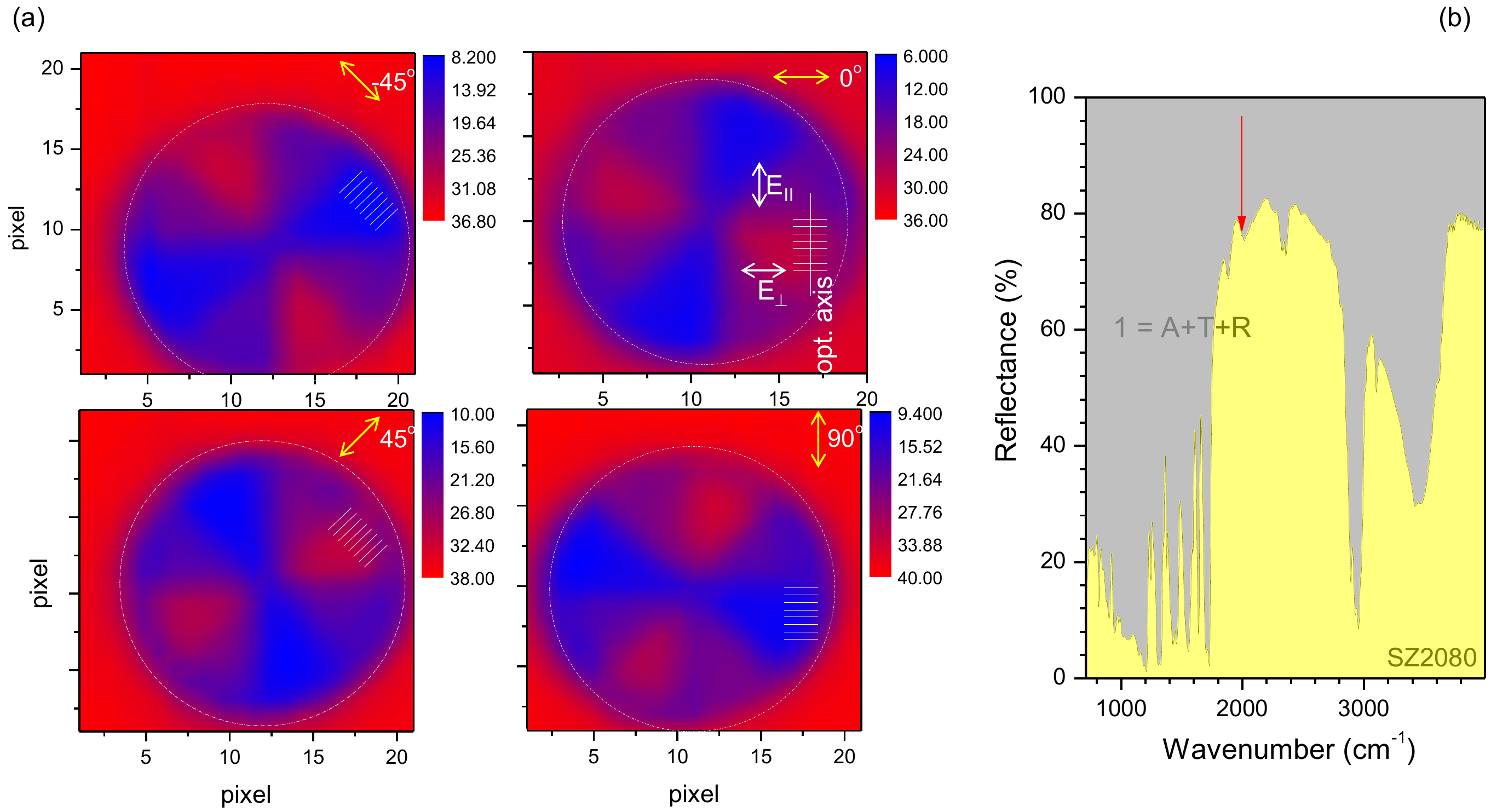}
\caption{Sub-wavelength case in reflection, $\Lambda < \lambda$
($0.6 < 5~\mu$m). (a) Reflectance map, $A$,  of $q = 1$ plate at
2000~cm$^{-1}$ (5~$\mu$m) wavelength measured at four
polarisations. Sample is shown on middle row, second from top in
Fig.~\ref{f-hight}(c). Grating orientation of $q = 1$ plate is
schematically shown; $\Lambda = 0.6~\mu$m. Orientation of
polarisation of the ordinary ($n_o$) $E_\perp$ and extraordinary
($n_e$) $E_\parallel$ beams are schematically shown ($n_o > n_e$).
Pixel size was $x\times y = 6\times 5~\mu$m$^2$ (PerkinEmler,
Spotlight). (b) Reflectance $R = 1-A-T$ spectrum of SZ2080 resist
(normalised to Au mirror). Arrow marks 5~$\mu$m wavelength used
for reflectance mapping shown in (a).} \label{f-ir}
\end{figure}

The manually measured retardance with the Berek compensator under
white light illumination was compared with an automatic equivalent
compensator based on light crystal compensator (Abrio) at 546~nm
wavelength. Figure~\ref{f-abrio} shows retardance and orientation
of slow-axis along the ordinary refractive index $n_o > n_e$ in
the form-birefringent negative uniaxial pattern of the two
selected q-plates. High quality and uniformity of the pattern is
confirmed for the height as well as orientation over the entire
area of the q-plates. In order to compare retardance measurements
with two different compensators Berek and Abrio, the $\lambda/4$
waveplates for 530~nm and 532~nm wavelength were carried out using
setups located in Tokyo and Bordeaux, respectively. For Berek
(Tokyo), retardation value was $147.3 \pm 3.8$~nm (10
measurements) under white light illumination. For Abrio
(Bordeaux), the measured retardance of 132.4~nm was $\sim 0.5\%$
smaller than the theoretical value for the 0-order 532~nm
waveplate (measured at 546~nm). The Berek measurements carried out
manually over entire white light spectrum were by $\sim 11\%$
larger than the theoretically expected value. The values
summarised in Table~\ref{table1}, which were measured with Berek
compensator, can have an approximately $10\%$ over-estimate, which
is acceptable for the manual spectrally broad-band measurement of
the retardance.

Figure~\ref{f-ir}(a) shows reflectance of q-plate at IR 5~$\mu$m
wavelength at four polarisations and (b) shows its spectrum. The
height of the gratings constitutes only $\sim\lambda/10$  at this
mid-IR wavelength (Fig.~\ref{f-high}). Dichroic losses are usually
measured in transmission for two linearly polarised beam
orientations $e^{-\Delta^{"}} = \sqrt{P_\parallel/P_\perp}$, where
$P_{\parallel,\perp}$ are the transmitted power parallel and
perpendicular to the local optical axis, respectively. It can be
applied for the reflectance as well, e.g., for the $0^\circ$
orientation (Fig.~\ref{f-ir}(a)) the $R_\parallel/R_\perp$ ratio
gives $\sqrt{R_\parallel/R_\perp} = \sqrt{6/36} = 0.41$ or
$\Delta^{"} = 0.9$~rad at $5~\mu$m wavelength. Following a general
definition $\Delta = \Delta^{'}+i\Delta^{"}$ with
$\Delta^{(',")}=k[n_{\parallel}^{(',")} - n_{\perp}^{(',")}]h$
being the phase retardance $\Delta^{'}$ and dichroism $\Delta^{"}$
for the $h$ height of q-plate along the light propagation length
at wavevector $k = 2\pi/\lambda$ and wavelength $\lambda$.

The anisotropy of absorbance, $A$, can be determined from the
angular dependence of $A_\theta$ and only four angles with angular
separation of $\pi/2$ are required to make the fit~\cite{Hikima}:
\begin{equation}\label{e-4a}
A_\theta = A_\perp\cos^2(\theta) + A_\parallel\sin^2(\theta),
\end{equation}
\noindent where $A_{\parallel,\perp}$ are the absorbances parallel
and perpendicular to the local optical axis, respectively,
$\theta$ is the azimuthal angle. For the complex refractive index
$n(\omega) = n(\omega)' + n(\omega)"$, the oscillating E-field of
light can be written as a function of height, $h$, as $E(h) =
E_0e^{i\omega(\frac{nh}{c}-t)} = E_0e^{-\omega n" h/c} \times
e^{i\omega(\frac{n' h}{c} -t)}$, where $\omega$ is the angular
frequency and $c$ is speed of light. The amplitude of the E-field
decreases exponentially with high, i.e., the intensity is given by
the Lambert-Beer's law $I(h) = I_0e^{-2n"\omega h/c} =
I_0e^{-\alpha(\omega)h}$, where $\alpha(\omega) = 2n"k$ is the
absorption coefficient. Then, the amplitude, $Amp$, of the
$\sin$-wave-form measured by the 4-polarisation method
(Eqn.~\ref{e-4a}) is related to the dichroism as:
\begin{equation}\label{e-link}
  Amp = (A_{\parallel} - A_{\perp})/2 = (\alpha_\parallel(\omega)h -
\alpha_\perp(\omega)h)/2 = k(n"_\parallel - n"_\perp)h \equiv
\Delta".
\end{equation}

\section{Discussion}

For the case when the period of structure is comparable to the
wavelength, $\Lambda\sim\lambda$, analytical methods to calculate
effective refractive index by EMT approach lose
validity~\cite{wolf,Emoto} and direct measurements of retardance
should be carried out. The measured retardance of flat optical
elements made of azimuthally oriented segments of gratings with
0.6, 1~$\mu$m period and a duty-cycle close to 0.5, shows that the
phase retardance can be controlled with high precision and a $\pi$
value can be made in reflection over the entire visible spectral
range. Taller polymerised structures would be required to make
$\lambda/2$ waveplates (with $\pi$ phase retardance) in
transmission, also, for longer IR-wavelengths. For telecom
spectral range around $\sim 1.5~\mu$m, taller structures could be
fabricated with the periods in sub-wavelength range (similar to
that used in this study). Higher aspect ratio 3D polymerised
structures are in reach for fs-laser polymerisation, especially,
when critical point dryer (CPD) is used to retrieve fabricated and
developed sample from a rinse solution. By avoiding capillary
forces in super-critical CO$_2$, it is possible to recover 3D
patter ns with intricate 100~nm feature sizes without mechanical
failure~\cite{16le16133}.

For q-plates with lower than $\pi$ retardance in phase there is a
possibility to separate the optical vortex beam which carries OAM
from the non-vortex part. Q-plates are irradiated with a
circularly polarised beam which carries a defined spin angular
momentum (SAM) and the vortex beam with (OAM) generated via
spin-orbital coupling in q-plate has a counter-circular SAM. By
separation of left and right circular polarised beams it is
possible to use vortex generation at a cost of lower efficiency.

The efficiency and purity of vortex generation are discussed next.
Transmittance for E-field $\tau = e^{-k(n_{\parallel}^{"} +
n_{\perp}^{"})h/2}$ accounts for the losses, hence, $\tau^2$
defines the overall transmittance; intensity $\propto E^2$. The
electric field emerging through the q-plate, $\mathbf{E}_{out}$,
is given~\cite{16aom306}:
\begin{equation}\label{e1}
\mathbf{E}_{out}= \mathbf{E}_{in}\tau
[\cos(\Delta/2)\mathbf{e}_\sigma + i\sin(\Delta/2)e^{i2\sigma
q\alpha}\mathbf{e}_{-\sigma}],
\end{equation}
\noindent where the incident field is circularly polarised
$\mathbf{E}_{in}\propto \mathbf{e}_{\sigma = \pm1}$ with
$\mathbf{e}_{\sigma} = \frac{1}{\sqrt{2}}(\mathbf{x} + i\sigma
\mathbf{y})$ and $\sigma = \pm 1$ defining the left and
right-handed polarisations in the Cartesian frame, respectively.
At the given transmittance, the spin-orbital coupling efficiency
is determined by the dichroism and birefringence properties of the
material and is maximised for the half-wavelplate condition of the
q-plate: $\Delta^{'} = \pi$ modulo $2\pi$.

The purity of the spin-orbital conversion is defined by parameter
$\eta$, which is the fraction of the output power that corresponds
to helicity-flipped field expressed as $\eta = \left(1 -
\cos\Delta'/\cosh\Delta''\right)/2$\cite{DavitPhD}. Interestingly,
the dichroism may enhance or reduce the purity depending on the
real birefringent phase retardation~\cite{17apl}. At the
half-waveplate condition $\Delta^{'} = \pi$, $\eta = 1$ when
dichroic losses are absent $\Delta^{"} = 0$. Dichroic losses are
measured for the $q=0$ plate (a grating) for two linearly
polarised beam orientations as shown above. The vortex generation
efficiency is $\tau^2\eta$.

\section{Conclusions and outlook}

High-quality sub-1~mm flat micro-optical elements based on a
geometrical phase realised by azimuthally orientated grating
segments down to resolution of $\sim 150$~nm are demonstrated.
Laser writing using only fs-laser oscillator took only 0.5~hours
to fabricate entire optical element and this is a simpler solution
compared with polymerisation of spiral plate~\cite{10apl211108}.
An increase of retardance using optical path doubling in
reflection mode allowed to achieve $\pi$ phase control required
for high purity of optical converters which turn circular
polarisation (a spin angular momentum) of an incoming light into
counter-circularly polarised optical vortex (a beam with the
orbital angular momentum). Such planar optical vortex generators
can find applications in optical manipulation, microfluidics, and
sensing~\cite{99pps665,09jnopm167,07oe12979}. Sub-wavelength
period gratings at telecommunication spectral window
1.3-1.5~$\mu$m can be made using the proposed method.

Other strategies to obtain reflective optical vortex generators
based on spin-orbit optical interactions using chiral and
anisotropic media have recently been introduced~\cite{e1,e2,j1}.

\acknowledgments 

We are grateful to Fujii-san from Nichika, Co. Ltd. for providing
us (within the same day of inquiry) with a calibration curve of
the Berek compensator No. 10412 which was purchased by Tokyo
Institute of Technology in around 1995. J.M. acknowledges partial
support by the Kakenhi No. 16K06768 grant, S.J. was partially
supported by the NATO grant SPS-985048 and the Australian Research
Council DP170100131 Discovery Projects. Research visit of S.J. to
Tokyo Institute of Technology was supported via the Australian
Academy of Science and JSPS fellowship scheme in 2016.
\bibliographystyle{spiebib} 

\end{document}